# How Many Bits Are Required for RIS Designs Without Far-Field Quantization Lobe?

Ruiqi Wang, Yiming Yang, and Atif Shamim
Electrical Engineering Department, King Abdullah University of Science and Technology (KAUST), Thuwal, Saudi Arabia,
*ruiqi.wang.1@kaust.edu.sa*

***Abstract*—Reconfigurable Intelligent Surface (RIS) designs with 1-bit phase resolution often suffer from strong quantization lobes in the far field, which significantly degrade wireless communication performance. This work investigates the minimum phase resolution required for RIS to eliminate far-field quantization lobe. The analysis demonstrates that 2-bit phase discretization offers an optimal balance between performance and hardware complexity. A practical 2-bit RIS unit cell is designed, and a 20×20 array configuration is implemented to evaluate its performance. The quantization-lobe suppression capability is validated through full-wave radar cross-section (RCS) simulations under plane-wave illumination for the entire RIS array. The fabricated prototype is further characterized experimentally, achieving a −13.1 dB quantization-lobe level compared to −0.8 dB for its 1-bit counterpart, confirming both the analytical and full-wave simulation results.**

***Index Terms*—Reconfigurable Intelligent Surface (RIS), wireless communication, phase quantization, quantization lobe, far field.**

## I. Introduction

The upcoming sixth-generation (6G) wireless systems aim to deliver extremely high data rates, massive connectivity, and ultra-low latency. However, in practical environments, electromagnetic (EM) waves are often blocked by obstacles, resulting in severe line-of-sight (LOS) limitations. To overcome this issue, Reconfigurable Intelligent Surfaces (RIS) have been proposed to create an alternative LOS link, thereby enhancing signal coverage and reliability [1], [2].

Unlike conventional reconfigurable reflectarrays, where the feed horn is typically placed in the near field of the surface, RIS are required to operate effectively in both near- and far-field regions [3]. When operating in the near field, a 1-bit RIS can still maintain a dominant main beam toward the desired direction despite elevated sidelobe levels. In contrast, in the far field, 1-bit RISs exhibit a severe quantization lobe [4], as illustrated in Fig. 1, whose strength can approach that of the main beam. Such strong undesired scattering significantly degrades communication quality [5]. Several techniques have been proposed to mitigate quantization lobes, such as introducing random phase delays [6]. However, these methods demand complex unit-cell designs and array configurations. Alternatively, varactor-based RISs can achieve higher phase resolution, yet they suffer from excessive loss at high frequencies [7]. To date, no prior work has systematically quantified the minimum number of phase bits required to suppress far-field quantization lobes, which is crucial for balancing performance and cost.

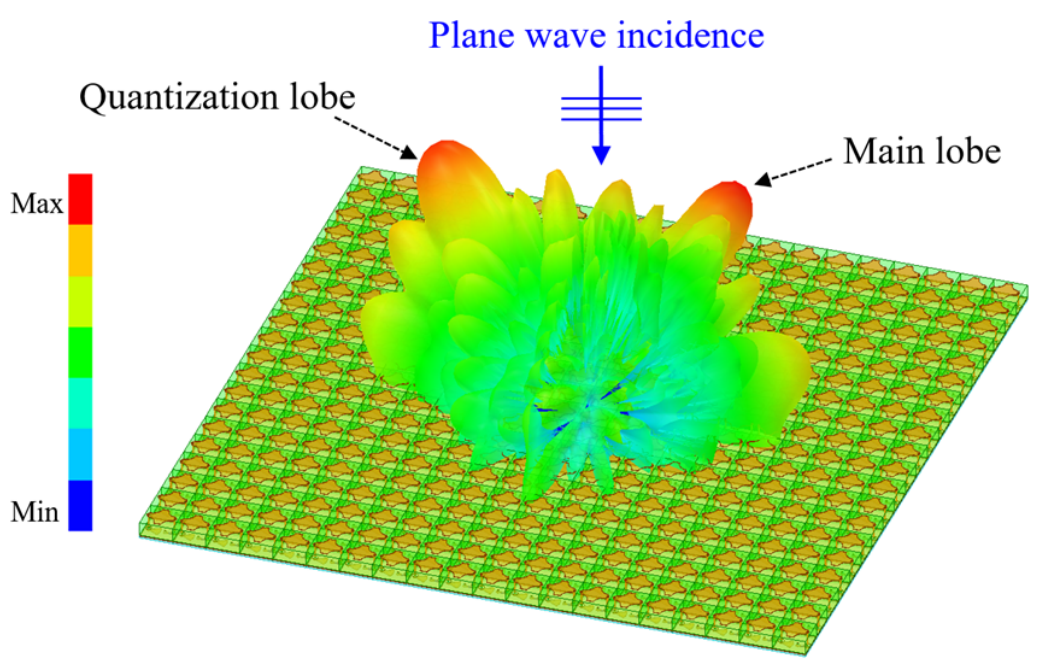

Fig. 1. Illustration of the far-field quantization lobe for the 1-bit RIS.

In this work, the relationship between phase resolution and far-field radar cross-section (RCS) performance is analyzed. Three representative cases, namely 1-bit, 2-bit, and 3-bit RISs, are compared with a continuously tunable surface. The results indicate that 2-bit phase quantization effectively suppresses quantization lobes while maintaining a practical implementation cost. A 2-bit RIS unit cell is then designed, and a 20 × 20 array is subsequently constructed, simulated, and experimentally evaluated. The measured results show that increasing the phase quantization level from 1-bit to 2-bit reduces the quantization-lobe level from −0.8 dB to −13.1 dB, thereby confirming the analytical predictions and full-wave simulation outcomes.

## II. Quantization Lobe Characterization

The far-field response of the RIS under normal incidence is first analyzed. A configuration with 0° incidence and 30° reflection is considered. The RIS consists of a 20 × 20 array with a unit-cell size of 4.3 mm, operating at 27.5 GHz, which is consistent with the experimental platform described in Section IV. The excitation source is placed 20 m away to ensure far-field illumination. The calculated phase distribution and the normalized array factor are presented in Fig. 2.

From the array factor results, a pronounced quantization lobe of −3.8 dB appears at −30° for the 1-bit case, which is comparable in magnitude to the main beam at 30°. When the quantization resolution increases to 2-bit, this lobe is effectively suppressed to −18.1 dB near −35°. Further increasing the resolution to 3-bit slightly improves the suppression to −19.5 dB. Therefore, 2-bit phase quantization is sufficient to eliminate significant quantization lobes under normal incidence, with minimal improvement obtained beyond this resolution.

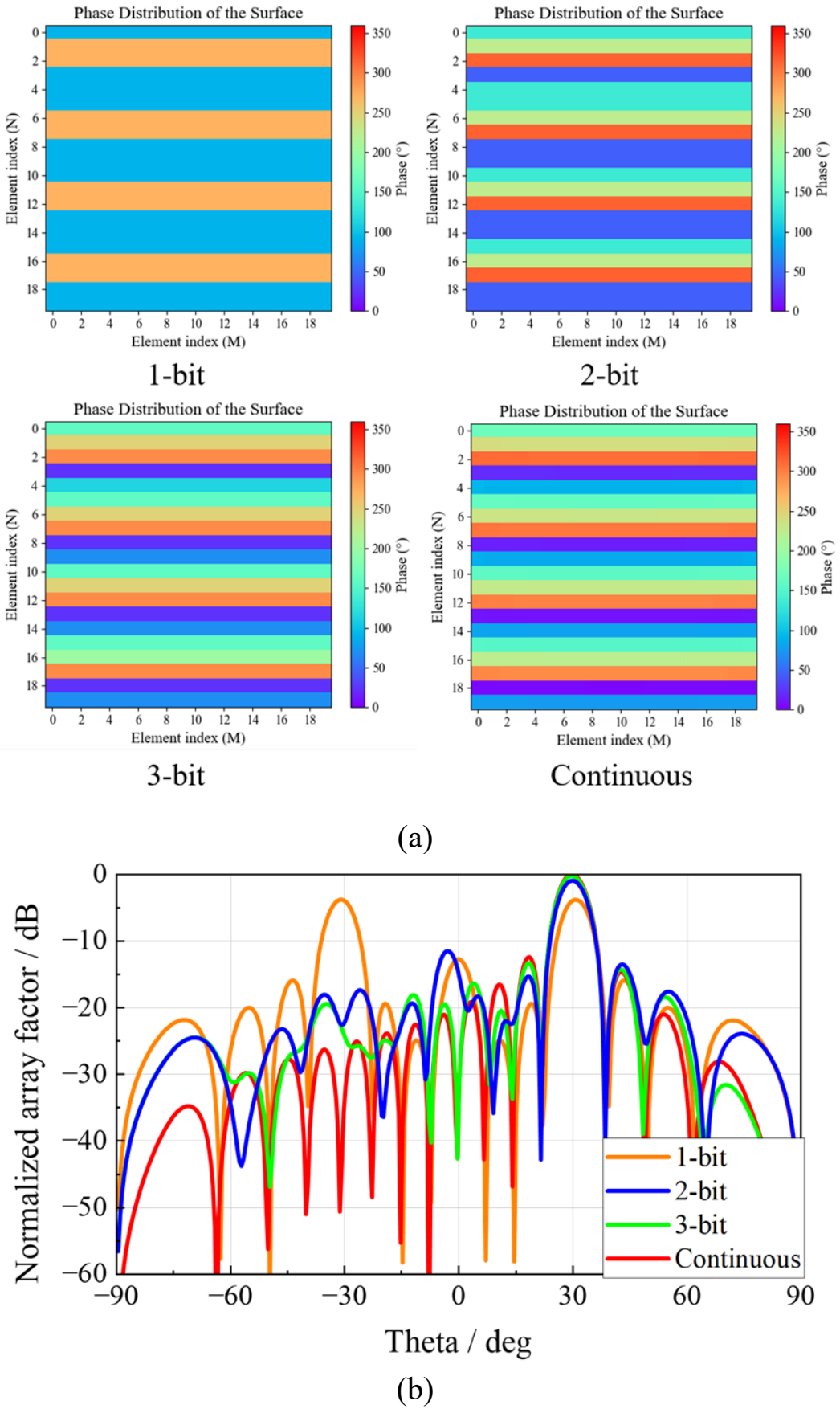

Fig. 2. Far-field characterization of the RIS with different phase quantization levels under normal incidence. (a) Phase distribution. (b) Normalized array factor.

Then, the effect of oblique incidence is analyzed by considering two scenarios: (i) 60° incidence with 0° reflection and (ii) 10° incidence with 50° reflection, as shown in Fig. 3. In the first case, a strong −8.0 dB quantization lobe appears at −52° in the 1-bit configuration and is suppressed to −18.0 dB in the 2-bit case. The 3-bit configuration offers no notable additional improvement. Similarly, for 10° incidence and 50° reflection, the 1-bit RIS exhibits a −0.9 dB quantization lobe at −24°, which is reduced to −20 dB when 2-bit quantization is employed. These observations confirm that 2-bit phase resolution provides a reliable and efficient solution for suppressing far-field quantization lobes under different incidence conditions.

## III. RIS Design and Simulation

To validate the above analysis, a practical 2-bit RIS unit cell is designed, as shown in Fig. 4. The structure consists of a top patch layer, an intermediate slot layer, and a bottom ground plane. Two PIN diodes are embedded in the slot layer, providing four reconfigurable states. By optimizing the slot lengths, a 90° phase shift between consecutive states is achieved. The patch and slot layers are fabricated on Rogers 5880 substrates, while the ground plane is implemented on an FR4 substrate to reduce cost.

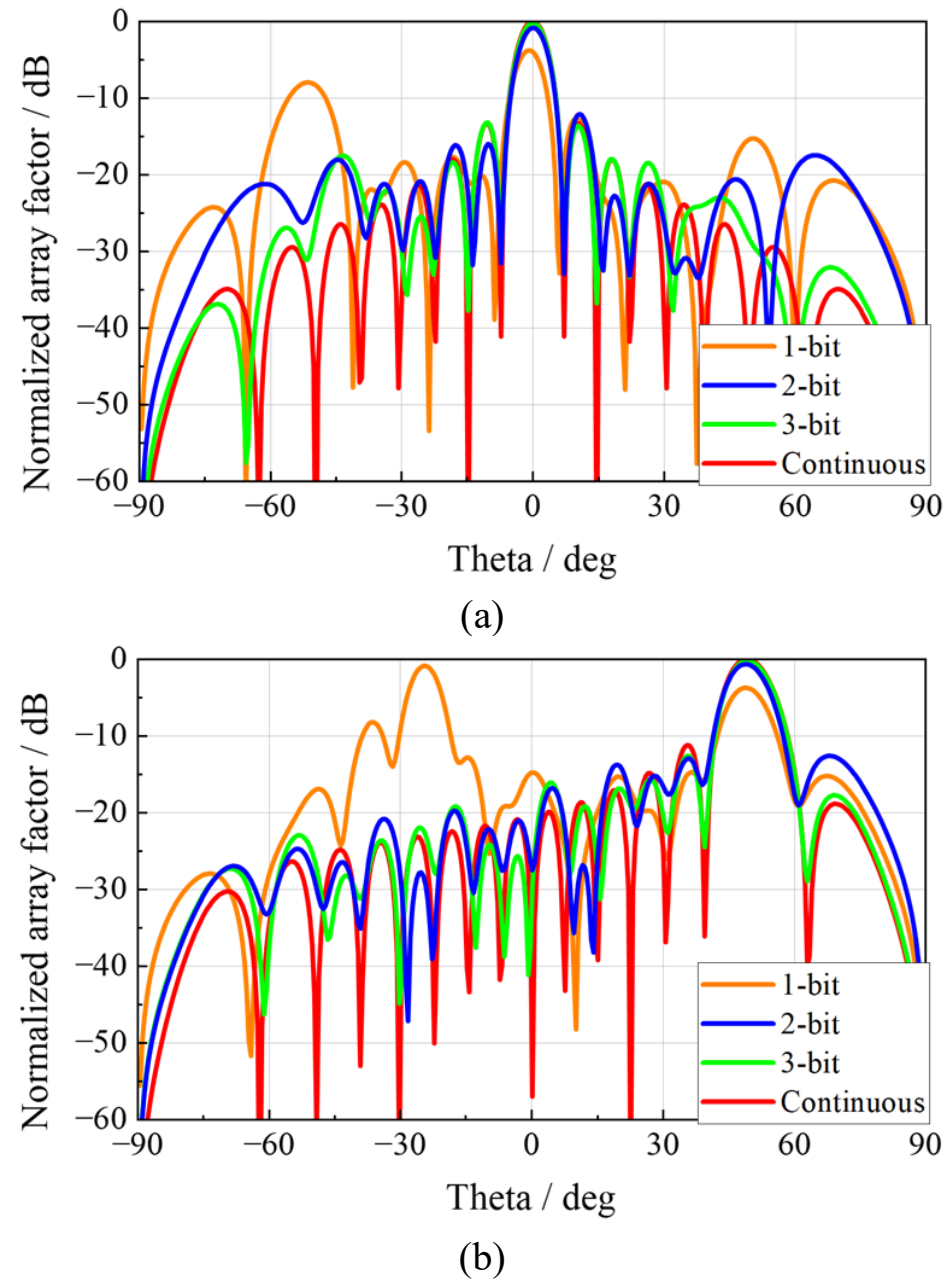

Fig. 3. Far-field characterization of the RIS with different phase quantization levels under oblique incidence. (a) 60° incidence and 0° reflection. (b) 10° incidence and 50° reflection.

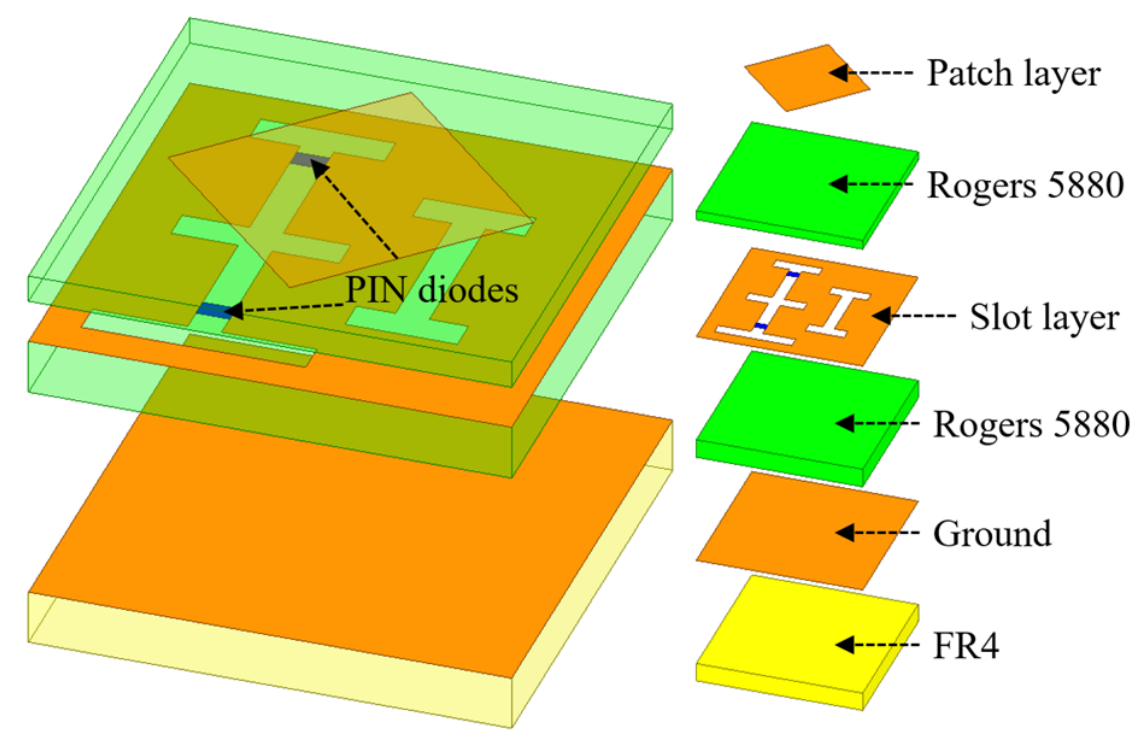

Fig. 4. Designed RIS unit cell with 2-bit phase resolution.

Then, a 20 × 20 RIS array is constructed using the designed unit cell to evaluate the far-field performance of the proposed configuration. The simulation setup, illustrated in Fig. 5, employs a normally incident plane wave and a target reflection angle of 30°, corresponding to the same condition analyzed in the theoretical model. Full-wave radar cross-section (RCS) simulations are carried out to assess the scattering characteristics of the entire array. As shown in Fig. 6, the quantization lobe is effectively suppressed in the 2-bit configuration compared with the 1-bit case in Fig. 1, demonstrating a significant improvement in far-field RCS response. Moreover, the simulated scattering pattern agrees well with the theoretical array-factor prediction in Fig. 2, validating the accuracy of the phase-discretization analysis and confirming the capability of 2-bit phase quantization to mitigate far-field quantization effects.

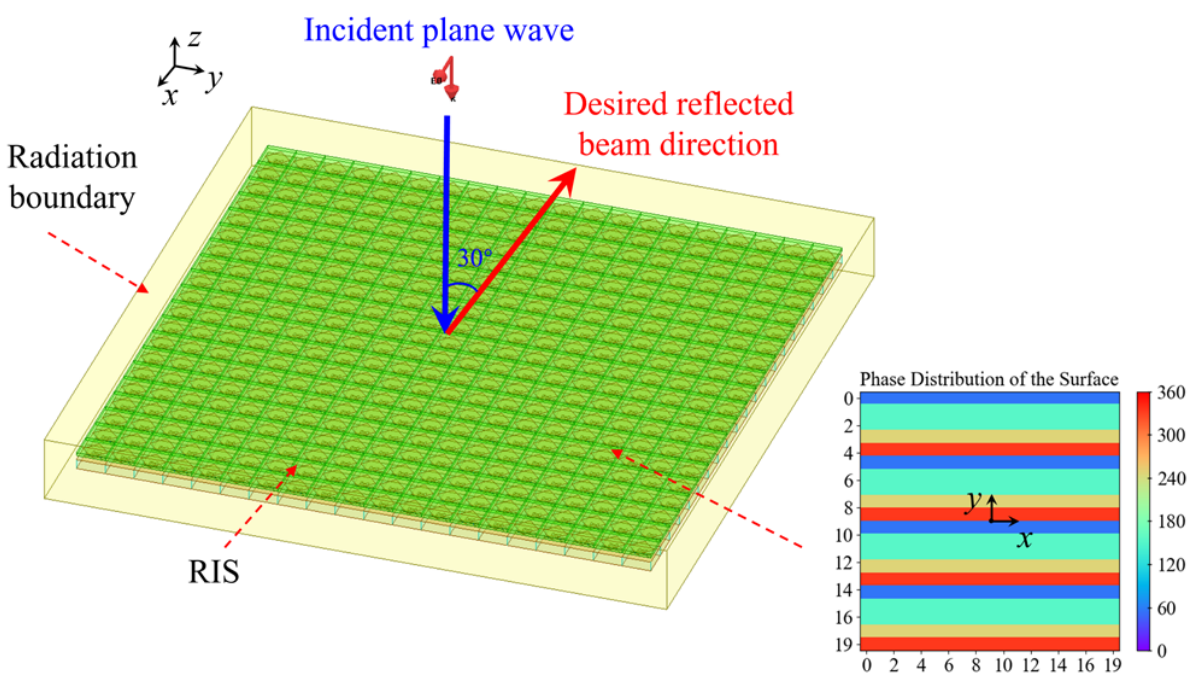

Fig. 5. Far-field RCS simulation of the RIS with 2-bit phase quantization.

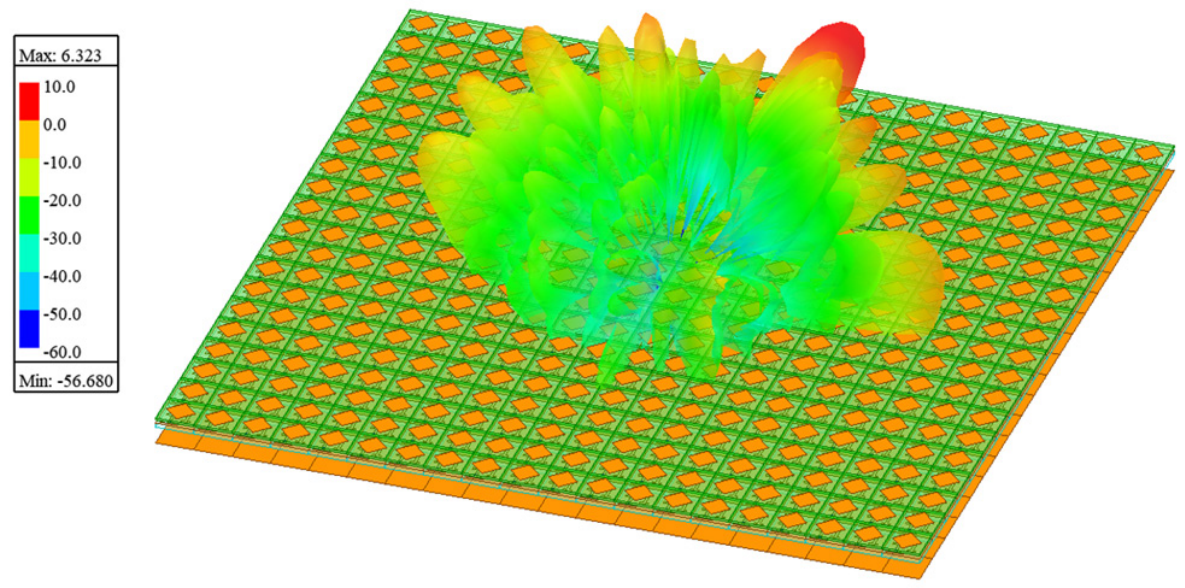

Fig. 6. Simulated far-field RCS results of the 2-bit RIS.

## IV. Experimental Validation

After the phase quantization analysis and full-wave simulations, the quantization-lobe suppression capability of the 2-bit RIS is further validated experimentally. The RIS design is fabricated, and the far-field measurement setup is illustrated in Fig. 7. To ensure consistency with the simulation configuration, the transmitting (Tx) horn antenna is positioned along the normal-incidence direction and connected to a signal generator operating at 27.5 GHz. The Tx antenna is located in the far-field region of the RIS. The receiving (Rx) horn antenna is placed at the desired reflection angle of 30° and connected to a spectrum analyzer to record the received power levels.

The RIS reflection pattern is configured under three conditions: the OFF state, 1-bit quantization, and 2-bit quantization. When the RIS is activated, the phase distribution corresponds to 0° incidence and 30° reflection. The measured received power levels for these three configurations are summarized in Table I. The results show that when the RIS is in the OFF state, the received powers at the main-lobe and quantization-lobe directions are nearly identical. When the RIS operates with 1-bit quantization, a 16.4 dB gain enhancement is observed at the main-lobe direction, indicating decent far-field beamforming performance. However, the quantization-lobe level remains close to the main beam, with only a 0.8 dB difference, consistent with the analytical prediction in Fig. 2 and the full-wave simulation in Fig. 1.

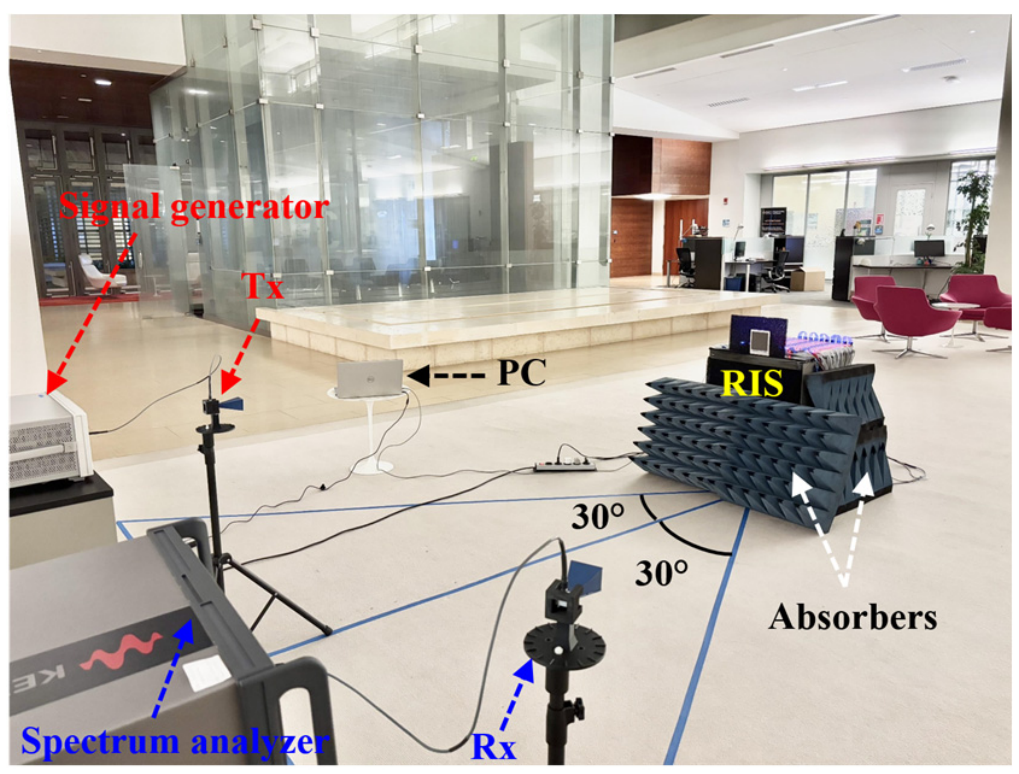

Fig. 7. Far-field experimental setup for the fabricated 2-bit RIS prototype.

TABLE I. Measured Received Power of the RIS in the Far Field

| Beam | Measured received power / dBm | | |
|---|---|---|---|
| | RIS OFF | RIS 1-bit ON | RIS 2-bit ON |
| Main lobe | -84.14 | -67.68 | -61.31 |
| Quantization lobe | -85.72 | -68.50 | -74.44 |

In contrast, when the phase resolution increases from 1-bit to 2-bit, the quantization lobe is significantly suppressed. A −13.1 dB difference between the main-lobe and quantization-lobe directions is achieved, confirming that 2-bit phase quantization effectively mitigates far-field quantization effect. These measurements experimentally validate the theoretical and simulated findings, demonstrating that 2-bit phase discretization offers a practical and efficient solution for RIS designs targeting far-field applications.

## V. Conclusion

In this work, the far-field quantization-lobe characteristics of reconfigurable intelligent surfaces (RISs) are systematically analyzed. RISs with 1-bit phase resolution suffer from pronounced quantization lobes that degrade far-field performance. To address this issue, the influence of phase-resolution enhancement is investigated. The results show that 2-bit phase quantization achieves an effective balance between lobe suppression and hardware complexity, whereas higher resolutions provide only marginal improvement. The conclusions are verified through full-wave simulations and further validated experimentally. The measurements confirm that a 2-bit RIS achieves a −13.1 dB quantization-lobe level, demonstrating the practical effectiveness of 2-bit phase discretization in suppressing far-field quantization artifacts and improving RIS performance.